\documentclass[english,showpacs,prb,superscriptaddress,twocolumn,amsmath,amssymb,floatfix]{revtex4-1}
\usepackage{natbib}
\usepackage{babel}
\usepackage{graphicx}
\usepackage{amsmath}

\newcommand{\kk}{{\bf k}}
\newcommand{\GG}{{\bf G}}

\newcommand{\qq}     {{\bf q}}
\newcommand{\pp} {{\bf p}}

\begin{document}

\title{Gate tunable quantum transport in double layer graphene}

\author{K. Kechedzhi, E.~H. Hwang, S. Das Sarma}

\affiliation{Condensed Matter Theory Center, Department of Physics, University
of Maryland College-Park, 20742-4111 MD USA}

\begin{abstract}
We analyze the effect of screening provided by the additional graphene layer in double layer graphene heterostructures (DLGs) on transport characteristics of DLG devices in the metallic regime. The effect of gate-tunable charge density in the additional layer is two-fold: it provides screening of the long-range potential of charged defects in the system, and screens out Coulomb interactions between charge carriers. We find that the efficiency of defect charge screening is strongly dependent on the concentration and location of defects within the DLG. In particular, only a moderate suppression of electron-hole puddles around the Dirac point induced by the high concentration of remote impurities in the silicon oxide substrate could be achieved. A stronger effect is found on the elastic relaxation rate due to charged defects resulting in mobility strongly dependent on the electron denisty in the additional layer of DLG. We find that the quantum interference correction to the resistivity of graphene is also strongly affected by screening in DLG. In particular, the dephasing rate is strongly suppressed by the additional screening that suppresses the amplitude of electron-electron interaction and reduces the diffusion time that electrons spend in proximity of each other. The latter effect combined with screening of elastic relaxation rates results in a peculiar gate tunable weak-localization magnetoresistance and quantum correction to resistivity. We propose suitable experiments to test our theory and discuss the possible relevance of our results to existing data.
\end{abstract}

\maketitle

\date{\today}

\section{introduction}

Gate tunable Anderson localization of Dirac electrons is inaccessible in standard field effect heterostructures due to the strong disorder-induced inhomogeneity of the local doping level (electron-hole puddles)~\cite{Adam2007,DasSarmaRMP11,SSETYacobyPuddles,hBNInhomogeneity,MLGonhBNrommie,WaterClusterDoping} ubiquitous in the materials that show Dirac-like low energy spectrum, such as graphene~\cite{MinCondGeim05} or topological insulator surfaces~\cite{KaneRMP10,*SCZhang11,*PuddlesTI}. Basically, the low-density localization behavior of graphene is inaccessible experimentally because the Dirac point itself is inaccessible due to the formation of electron-hole puddles around the charge neutrality point. However, very recently, a novel double layer graphene (DLG) heterostructure containing two graphene layers separated by an insulator allowed using additional screening effect of the second graphene layer to access a metal-insulator transition regime. Anomalously large resistance, $\rho \gg h/(4e^2)$, in combination with insulating temperature dependence was observed in these experiments suggestive of a metal insulator transition (MIT)~\cite{Geim11}. The physical nature of this behavior is currently a topic of a debate~\cite{Geim11,DisorderbyOrder}. In particular, whether the transport data of Ref.~\onlinecite{Geim11} represent an effective low-density high-temperature semi-classical resistivity~\cite{DisorderbyOrder} or a low temperature strong Anderson localization crossover behavior~\cite{Geim11} is unclear at this stage.
 
On the one hand suppressing inhomogeneity allows access to the low-density regime $k_F\ell\lesssim1$ in which Anderson physics is expected to dominate (here $k_F$ is the Fermi wavevector and $\ell$ is the mean free path). In this regime, quantum interference of the two flavors (due to two valleys) of chiral Dirac charge carriers in graphene may result in the insulating behavior in the case of sufficient mixing of the flavors by atomic-scale disorder~\cite{Geim11,AleinerEfetov06,*Altland06,*Ostrovsky07,*OstrovskyEPJ,NomuraMacDonald,*Bardarson07,*DelocDiracNumerics,*AndersonMLGNumerics,*AndersonTBMLGNum,*AndersonNumRoche,*AndersonRocheChemModMLG}. On the other hand, the experiments of Ref.~\onlinecite{Geim11} were restricted to a fairly high temperature regime (10-100K) where quantum interference effects may be suppressed due to short phase breaking length. The observed MIT is also in contrast with more recent transport measurements on ultra high quality suspended devices~\cite{SuspGeim12} in which an extremely low density inhomogeneity is achieved, $\delta n \sim 10^8cm^{-2}$, nevertheless no MIT is detected (see also more recent experiment Ref.~\onlinecite{StanfordTopGate}). Moreover, an Anderson insulator would be characterized by the resistivity growing exponentially with decreasing temperature whereas the data~\cite{Geim11} demonstrates only a roughly power law growth. Insulating power law temperature dependence (resistivity growing with decreasing temperature) is not unusual near charge neutrality in graphene (see Fig.~2 in Ref.~\onlinecite{Geim11}, and measurements on other low~\cite{RhoTKim2007,RhoTCVD} and high~\cite{SuspGeim12,SuspRhoTKim08,*SuspRhoTAndrei08} mobility samples). This behavior is explained by the combined effect of temperature dependent screening~\cite{HwangScreenRhoT} thermal excitation of electrons from the valence band and thermally activated hopping of electrons over the potential barriers between electron-hole puddles~\cite{BoltzmanInsulatingTdep}. Nevertheless, very high values of resistance reported in Ref.~\onlinecite{Geim11}, $\rho \gg h/(4e^2)$ suggest a novel behavior in this system possibly associated with MIT. However, an alternative explanation for the observed behavior was recently suggested~\cite{DisorderbyOrder,BoltzmanInsulatingTdep}, in which case the anomalous resistance is explained by a Boltzmann transport effect combined with strong suppression of inhomogeneity in DLG~\cite{DisorderbyOrder,BoltzmanInsulatingTdep}. The latter work is phenomenological and relies on the assumptions of the quasiclassical transport formalism and the phase breaking length being shorter than the elastic scattering length. Therefore a more detailed analysis of quantum effects is required to justify the applicability of the latter approach. One way to make progress in the understanding of the observed behavior of the resistivity in these new devices is to analyze the metallic regime where perturbation expansion in disorder strength may be applied. Such theoretical analysis of the quantum interference effect in the higher-density weak localization regime of the DLG system, which must be a precursor to any low-density strong localization crossover phenomenon, is one of the main goals of our work.

In this paper we consider the effect of gate tunable screening provided by the additional layer in DLGs on transport characteristics of these devices. Screening affects both the potential of charged defects and Coulomb interactions of charge carriers. Taking into account both effects we analyze the classical and quantum parts of the resistivity of graphene in DLG geometry in the metallic regime $k_F\ell \gg 1$. We consider only low temperature regime such that elastic scattering by the disorder limits the transport characteristics and all inelastic scattering effects may be included perturbatively. 

Below we outline the results of the following analysis: (i) We develop a framework describing the screening effect of the two graphene layers in DLG on the Coulomb potential of charged defects in DLG heterostructure. (ii) We generalize the self-consistent theory of disorder induced electron density fluctuations at neutrality point of graphene~\cite{Adam2007} (electron-hole puddles) and use it to estimate the screening effect of DLG on the amplitude of the electron density fluctuations. (iii) We analyze the effect of screening in DLG on the elastic scattering rate due to charged defects that determines the extent of gate tunability of mobility in DLG. (iv) We analyze the effect of the screening in DLG on the dephasing rate due to inelastic electron-electron collisions. (v) We analyze the combined effect of screening on the weak localization correction to the resistivity (vi) We predict a peculiar gate tunable magneto-resistance in the high quality DLG structures with mixed elastic scattering mechanisms. 

The present work is limited to the perturbative metallic regime and cannot make any definitive conclusions about the regime of the low density studied in the experiments of Ref.~\onlinecite{Geim11}. However, the theoretical framework developed below provides a basis for analysis of a more detailed gate tunable quantum transport measurements which would allow characterization of various microscopic scattering mechanisms that determine the behavior of the system in the low density regime. In particular, we believe that it is imperative that we understand the high-density metallic regime of the DLG transport in some depth before trying to understand the low-density crossover MIT regime since a systematic perturbative theory is available in the higher density regime whereas the low-density regime is inaccessible to analytical theory. We hope that our work would motivate experimental work focusing on the metallic regime of DLG in order to investigate conceptually important question of Anderson localization and metal-insulator transition in graphene.

\begin{figure}[t]
\includegraphics[width=0.25\textwidth]{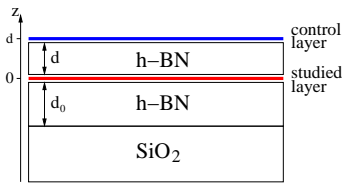}
\caption{(color online) A schematic of a typical double layer graphene heterostructure used in Ref.~\onlinecite{Geim11}. Top (blue) and bottom (red) layers correspond to the ``control'' and ``studied'' layers respectively with typical values of $d=4nm$ and $d_0=20nm$ considered in the text. }\label{fig0:sketch}
\end{figure}

We consider a DLG consisting of two layers of graphene, ``studied'' and ``control'', see Fig.~\ref{fig0:sketch}, which are used for resistivity measurements and to provide the additional screening effect, respectively. The two layers are separated by a thin layer, $d\gtrsim 4nm$, of insulating hexagonal boron nitride (hBN) that is thick enough to suppress direct tunneling between the layers. The whole structure is separated by a thick layer, $d_0\sim 20nm$, of hBN from the standard SiO$_2$/Si substrate. The combination of the top and bottom gates in this heterostructure allows independent ``control'' of the electron densities $n_c$ and $n_s$ in the ``control'' and ``studied'' layers respectively. This setup is motivated by the design of heterostructures in Ref.~\onlinecite{Geim11}. We assume that the mobility of the ``studied'' layer is limited by elastic scattering of charged impurities in accordance with the data~\cite{Geim11}.

The additional screening effect of the ``control'' layer is expected to suppress the electron-hole puddles in the ``studied'' layer. We find that the effectiveness of the ``control'' layer screening strongly depends on the position of impurities in the DLG. It is efficient for the relatively small concentration of charged defects on top of the ``control'' layer, whereas it leads to only a moderate suppression of the inhomogeneity induced by charges in close proximity of the ``studied'' layer and by the remote charged defects in SiO$_2$ substrate. The latter inefficiency stems from the very high densities of charged defects in the SiO$_2$ substrate, $n_i\gtrsim 10^{12}cm^{-2}$, that would require screening charge density in the ``control'' layer several times $n_i$ which is beyond the experimentally accessible range. Moreover, the nonlinear screening effects in the electron-hole puddle regime in the ``studied'' layer work against the screening effect of the ``control'' layer making the screening less efficient. The situation is slightly different in the case of elastic relaxation rate due to charged defects. The screening by ``control'' layer is more efficient in this case and leads to the suppression of the classical part of the resistivity near the Dirac point of the ``studied'' layer. The resulting behavior of the resistivity as a function of the density in the ``control'' layer is a result of a competition of the suppression of electron-hole puddles that increases the resistivity and suppression of the elastic relaxation rate that decreases the resistivity. 

Below we also demonstrate that the ratio of momentum relaxation rate to the transport rate varies substantially with the distance between graphene and charged defects. Therefore measurement of the two elastic relaxation rates can be used to determine the location of charged defects limiting the mobility in DLG. Transport scattering rate can be extracted from Boltzmann resistivity. The momentum relaxation rate characterizes the broadening of single particle states and can be extracted from the decay of Shubnikov de Haas oscillations. A similar technique~\cite{MomeRelHwang} has been successfully applied to the analysis of the standard graphene based field effect transistors~\cite{MomRelPenn09}. Since our work establishes that the inhomogeneous puddles induced by the distant charged impurities in SiO$_2$ can be suppressed only partially by the control layer screening, it is important to know where the dominant charged impurities reside and the measurement of the single-particle momentum relaxation (sometimes also called the quantum scattering) time could help locate the charged impurities in the DLG system.

Very high values of the mobility of the ``studied'' layer encapsulated in hBN that is tunable in the range $50-120\times10^3cm^2V^{-1}s^{-1} $ with the screening density in ``control'' layer $n_c$ indicate the dominant role of the charged defects in the elastic scattering in this system. Furthermore, the very high value of the observed~\cite{Geim11} maximum of the gate tunable mobility suggests a weak effect of neutral (immune to screening) elastic scattering mechanisms that break locally the lattice symmetry of graphene. Therefore these DLGs may demonstrate quantum interference behavior distinct from that observed in typical graphene flakes~\cite{NadyaMason,Savchenko08,KoreanGroup,SPMImpurity,SavchenkoWLWAL,WLvEEI,EEIEpitaxial,BrunSchweigEpitax,deHeer,Korean2010} where strong lattice symmetry breaking disorder determines predominantly weak-localization behavior. This is especially so at relatively low densities $n_s$ at which the effect of trigonal warping of the Fermi line~\cite{Kechedzhi06} can be neglected. 

Two identical carbon atoms in the unit cell of graphene give rise to two degenerate flavors (valleys) of chiral Dirac charge carriers in its electronic spectrum confined to the vicinities of the two inequivalent corners of the hexagonal Brillouin zone. Quantum interference of two independent flavors of Dirac quasiparticles is associated with weak anti-localization behavior~\cite{Ando02,*Khvesh06,*MorpugoGuinea,Kechedzhi06}. The latter is however protected only by the symmetry between the two sublattices in graphene, which can be easily broken by impurities. Therefore quantum interference in graphene is very sensitive to the presence of elastic scattering mechanisms that break lattice symmetry. Generic time inversion symmetric disorder in graphene can be categorized into three types according to their effect on quantum interference: (I) potential of the charged defects that does not break lattice symmetry; (II) smooth random vector potential disorder that breaks the symmetry between the two carbon atoms in the unit cell of graphene (intravalley disorder); (III) atomic scale disorder that breaks lattice symmetry and causes mixing of the two valleys (intervalley disorder). Random vector potential disorder leads to suppression of the weak-antilocalization effect whereas the intervalley scattering restores the weak localization behavior typical for two dimensional electron gas. In the case of high quality DLG heterostructures with very weak lattice symmetry breaking scattering weak antilocalization effect of Dirac electrons plays an important role. 

Additional screening in DLG affects only the scattering rate due to charged defects in graphene and therefore changes the relative strength of intra and inter valley defects with respect to the strength of potential scatterers. This results in the dependence of quantum correction on the electron density in the ``control'' layer $n_c$. In particular, an interesting possibility arises that the intervalley elastic scattering, while being unimportant in determining the semiclassical Boltzmann resistivity, becomes important in determining the tuning of the quantum interference correction from anti-localization to localization.

In a wide temperature range quantum interference is limited by decoherence rate due to inelastic electron-electron collisions.  The latter is strongly affected by the screening effect in DLG for two reasons: due to suppression of the Coulomb interaction between charge carriers, and due to enhancement of the diffusion coefficient that reduces the time electrons spend in proximity of each other where interaction is the strongest. The combination of these two effects results in strong variation of the decoherence rate with the screening density. This gives rise to a peculiar gate tunable quantum correction to the resistivity of graphene in DLG devices which may change sign depending on the screening density in the ``control'' layer especially at higher temperatures. It should be emphasized that this gate-tunable quantum correction to the Boltzmann resistivity is manifest only when the phase breaking length is much larger than the elastic mean free path, i.e. at lower temperatures, so that the quantum interference is operational. Since our theory treats the quantum correction perturbatively, it applies only in the metallic regime at densities well above the crossover to strong Anderson localization.

The most striking consequence of the gate tunable quantum correction can be observed in the intermediate density regime in which the transport in the ``studied'' layer is metallic, $k_F\ell\gg1$, whence the Fermi energy is low enough so that the trigonal warping effect is negligible. The quantum correction in this case is strongly dependent on the electron density in the ``control'' layer. At sufficiently strong dephasing rate (sufficiently high temperature) this results in a gate tunable crossover from weak-antilocalizaiton type magnetoresistance to weak-localization type. A similar crossover that was temperature driven was demonstrated in graphene previously~\cite{deHeer,SavchenkoWLWAL}, our work demonstrates a similar tunability with the gate voltage in DLG systems.

Quantum part of the resistivity also includes Altshuler-Aronov correction to the conductivity due to the electron-electron interactions~\cite{AAbook,AkkermansBook} which typically has the sign enhancing the localizaiton effect and is expected to be suppressed by the additional screening effect which typically reduces the interaction strength. This does not have any effect on low field magnetoresistance which allows to study the quantum interference (weak localization) correction separately.

The paper is organized as follows. In section II we analyze the effect of ``control'' layer screening on Coulomb potentials of the charged defects. Sections III and IV discuss classical and quantum parts of the resistivity respectively. We provide a discussion of the relevance of our work to experiments, particularly the data of Ref.~\onlinecite{Geim11} where the DLG system was studied experimentally in Section V. We conclude in Section VI.

\section{Screening effect of the ``control'' layer}

\subsection{Linear screening in DLG}

We start with a linear screening model for DLG heterostructure. For simplicity we assume that a density $n_{i}$ of charged defects is located in a plane a distance $z$ from the ``studied'' layer (at $z=0$), which can be both above $z>0$ and below $z<0$ it. The ``control'' layer is located at a distance $d$ above the ``studied'' layer.   

We include both intra and inter layer Coulomb interactions between electrons in DLG described by a symmetric matrix $U_{\ell\ell'}$ with diagonal, $U_{11}=U_{22}=\upsilon_q$, and off diagonal elements $U_{12}=U_{21}=\upsilon_q e^{-qd}$ with $\upsilon_q=2\pi e^2/q$ being the 2D Fourier transform of the Coulomb potential. Here $\ell,\ell'=1,2$ correspond to the ``studied'' and ``control'' layers respectively. We calculate the interaction energy of an impurity charge $Z_1$ and a conduction electron charge $Z_2$ in the ``studied'' layer in the presence of the screening effect of electron gases in both ``studied'' and ``control'' layers. This can be found using linked cluster expansion of the thermodynamic potential~\cite{MahanBook} of the DLG. In the resulting infinite series we keep only the term proportional to the product of the two charges that represents the linear screening of the interaction potential,
	\begin{equation} 
	\Delta\Omega(\mathbf{R_1}-\mathbf{R_2})=Z_1Z_2\int\frac{d\mathbf{q}}{\left(2\pi\right)^{2}}\frac{\upsilon_qe^{-q|z|}}{\varepsilon_z(\mathbf{q})}e^{i\mathbf{q\cdot\left(R_{1}-R_{2}\right)}}, \label{ThermPot}	
	\end{equation}
where $q,\mathbf{R}_i$  are 2D wave vector and position vector, respectively, and $Z_1Z_2e^{-q|z|}\upsilon_q$ is the Fourier transform of the bare Coulomb interaction between the charges $Z_1$ and $Z_2$. For electrons we can put $Z_1=Z_2=1$. In Eq.~(\ref{ThermPot}) we introduced a dielectric function, 
	\begin{equation}
	\frac{1}{\varepsilon_z(\mathbf{q})}=1-\frac{e^{q|z|}\upsilon_{q}}{2}\sum_{\ell\ell'}D_{\ell\ell'}(q)\chi_{\ell\ell'}(\mathbf{q}).\label{dielvdens}
	\end{equation}
where the matrix $D_{\ell\ell'} (q)$ takes into account the dependence of Coulomb interaction energy on the spatial separation between the charges in $z$ direction,
	\begin{eqnarray*}
	D_{\ell\ell'} (q) = \left[
		\begin{array}{cc}
		2e^{-q|z|} 					     & e^{-q|d-z|} + e^{-q(d+|z|)} \\
		e^{-q|d-z|} + e^{-q(d+|z|)} & 2e^{-qd} \\
		\end{array}
	\right]. 
	\end{eqnarray*}
Here $\chi_{\ell\ell'}=\langle T_{\tau}\rho_{\ell}(\mathbf{q},\tau)\rho_{\ell'}(-\mathbf{q},0)\rangle$ is the density-density correlator which is a matrix in the layer index, $\ell=1,2$, including both intra and inter layer terms. The the density-density correlator is renormalized by the Coulomb interaction matrix which is given by an infinite perturbative series~\footnote{Coulomb interaction for graphene encapsulated in hexagonal boron nitride is characterized by a dimensionless constant $e^2/(\Xi\hbar v_F)\approx0.4$, where we took the dielectric constant to be $\Xi\approx5$.}. Within the Random Phase Approximation the result of the resummation of the perturbation series satisfies a matrix Dyson equation~\cite{DasSarmaDoubleWell,GraphiteScreening86},
	\begin{equation}
	\hat{\chi}^{-1}=\left(\hat{\chi}^{(0)}\right)^{-1}+\hat{U}.\label{eq:DysMat}
	\end{equation}
where $\hat{\chi}^{(0)}$ is the diagonal matrix of free particle polarization operators in the two graphene layers. Solving the Dyson equation~(\ref{eq:DysMat}) we arrive at the dielectric function in the form,
	\begin{gather}
	\frac{1}{\varepsilon_z(\mathbf{q})}=\frac{1+\upsilon_{q}\chi_{22}^{(0)}\left(1-e^{-qx}\right)}{d(q)},\label{dielfunc} \\
	d(q)\equiv \left(1+\upsilon_{q}\chi_{11}^{(0)}\right)\left(1+\upsilon_{q}\chi_{22}^{(0)}\right)-\upsilon_{q}^{2}e^{-2qd}\chi_{11}^{(0)}\chi_{22}^{(0)}, \nonumber
	\end{gather}
where $x\equiv |z-d|-|z|+d$. The dielectric function~(\ref{dielfunc}) demonstrates a strong dependence on both the impurity location $z$ and the interlayer distance $d$, and in the limit $d\rightarrow \infty$ approaches the monolayer form $1/\varepsilon(q)=1/(1+\upsilon_q \chi^{(0)}_{11})$.

Static screening is described by the zero frequency limit of the free particle polarization operator~\cite{Guinea06,HwangDasSarma2007},
	\begin{gather*}
	\chi^{(0)}_{ii}=g\nu_{i}\left(\theta\left(2k_F^{(i)}-q\right) +\theta\left(q-2k_F^{(i)}\right)\mathcal{F}(q) \right), \\
	\mathcal{F}(q) = 1 +\frac{\pi q}{8k_F^{(i)}}-\frac{1}{2}\sqrt{1-\frac{4\left(k_F^{(i)}\right)^2}{q^2}}-\frac{q}{4k_F^{(i)}}\arcsin \frac{2k_F^{(i)}}{q},  
	\end{gather*}
where $\theta(t)$ stands for Heaviside step function, $g=4$ stands for spin and valley degeneracy combined, $k_F^{(i)}$ and $\nu_{i}$ stand for the Fermi wave vector and the density of states per spin per valley at the Fermi level in the $i$-th layer, respectively.

\subsection{Non-linear screening in the electron-hole puddle regime}~\label{sec:e-h}

\begin{figure}[t]
\includegraphics[width=0.235\textwidth]{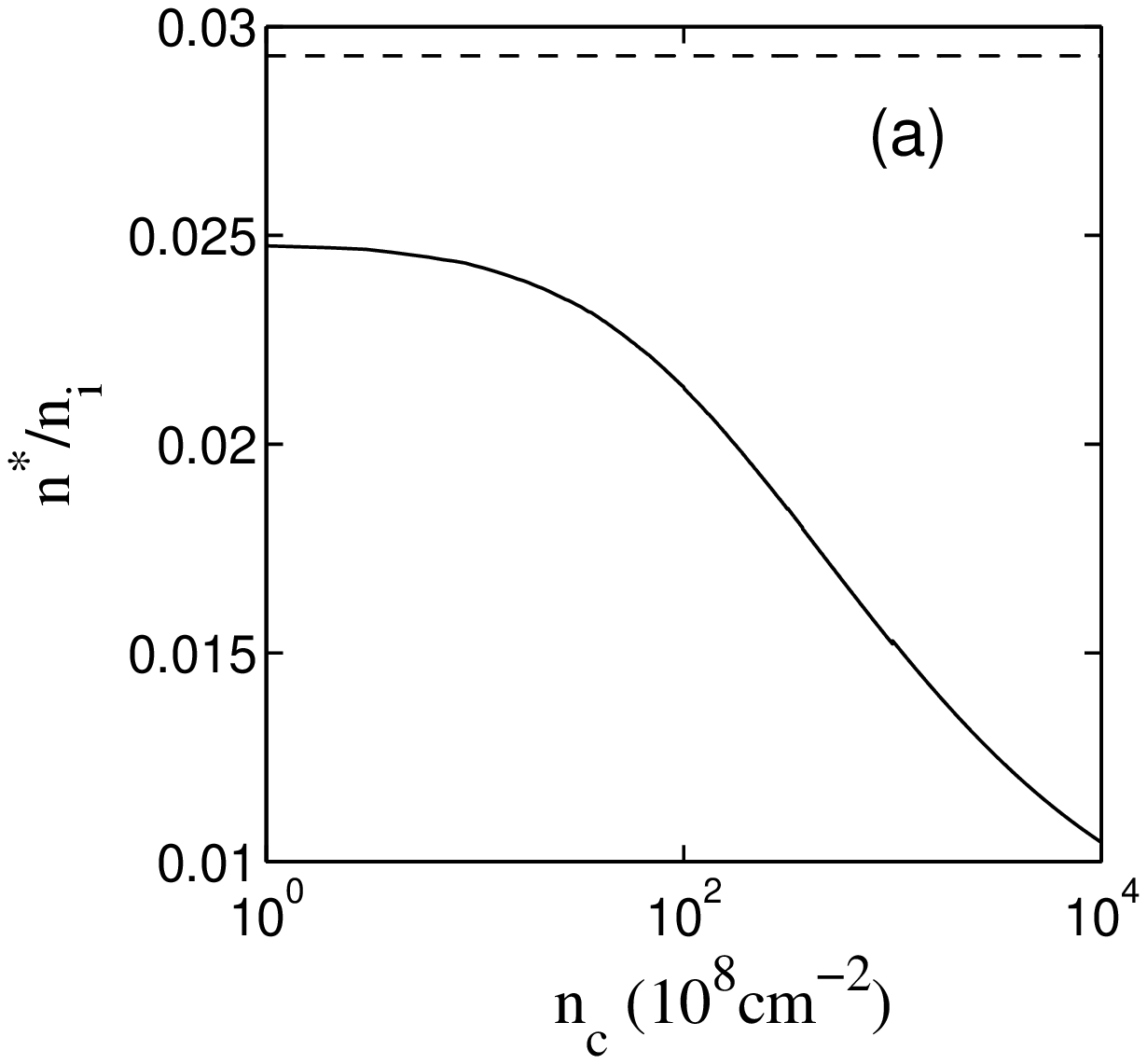}
\includegraphics[width=0.235\textwidth]{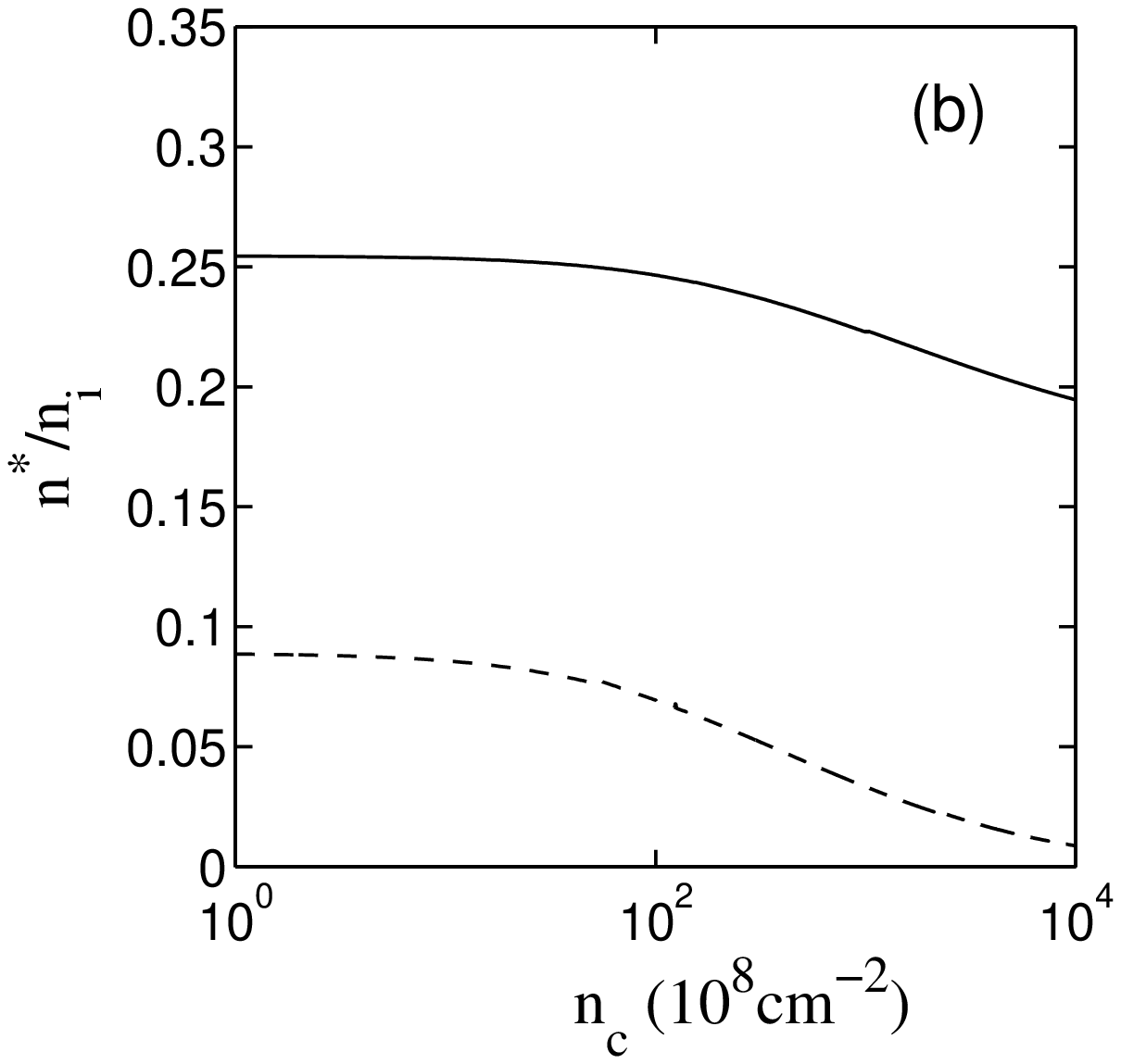}
\caption{ The result of the self-consistent density calculation in the ``studied'' layer
as a function of the density in the ``control'' layer $n_{c}$ for a DLG with $d=4nm$, see section~\ref{sec:e-h} for details. (a) Solid line corresponds to $z=-20nm$, $n_i=10^{12}cm^{-2}$. The horizontal dashed line corresponds to the result of the self-consistent 
density calculation for isolated monolayer graphene encapsulated in hBN with the same parameters as the solid line. (b) Solid and dashed lines correspond to $z=1nm$ and $z=5nm$  respectively with the impurity density $n_i=10^{11}cm^{-2}$. }\label{fig1:rmsns}
\end{figure}

We estimate the effect of screening by the ``control'' layer on the density inhomogeneity in the ``studied'' layer using the simple self-consistent approach that had shown good agreement with more detailed Thomas-Fermi-Dirac numerical calculations in the case of monolayer graphene~\cite{Galitski07,Adam2007,AdamRossi09,*Rossi08,*RossiEMT}. 
This estimate is based on the approximate relation $\langle E_F^2 \rangle \approx \langle V^2\rangle$, where the amplitude of the fluctuation of the local Fermi energy is equated to the local fluctuation in the potential energy of electrons, which reflects the local electrochemical equilibrium condition. We further assume that the local Fermi level is given by the free particle form $E_F=\hbar v_F k_F^{(1)} = \hbar v_F \sqrt{\pi n_s}$, where $v_F$ and $n_s$ stands for the Fermi velocity of Dirac electrons and local density respectively. We also assume that the screening of impurity potential can be taken into account within linear approximation~(\ref{dielfunc}). We then end up with a self-consistent equation for the typical density fluctuation $n^\ast$. The solution of this self-consistent problem is plotted in Fig.~\ref{fig1:rmsns} as a function of the ``control'' layer density $n_c$. The discrepancy between the solid black line in Fig.~\ref{fig1:rmsns}(a) as $n_c\rightarrow 0$ and the root mean square density in monolayer without the screening layer under same conditions, dashed line, originates from the screening by interband transitions in the ``control'' layer at neutrality. The above analysis neglects non-linear screening effects in the ``control'' layer and therefore the saturation in the realistic DLG device may be at a slightly different value. Nevertheless Fig.~\ref{fig1:rmsns} and the comparison to the monolayer case gives a reasonable estimate of the suppression of the density inhomogeneity by screening.

\subsection{Location of impurities in DLG}\label{sec:locImp}

Effectiveness of screening by the ``control'' layer is strongly dependent on the location of impurities ($z$) which can accumulate in three possible areas: Relatively small densities of charged defects, $n_i\sim10^{10}-10^{11}cm^{-2}$, can be located at hBN-graphene interfaces~\cite{DasSarmaHwanghBN} and on top of the ``control'' layer which can be exposed to air in realistic DLGs. A much larger concentration of charged impurities, $n_i\sim 10^{12}cm^{-2}$, is expected to accumulate at the surface of SiO$_2$ substrate that is separated from the ``studied'' layer by $d_0\sim20$nm of hBN, see Fig.~\ref{fig0:sketch}. These defects do not play an important role in transport in the high density regime $k_F|z| \gtrsim 1$ due to the exponential factors in Eqs.~(\ref{ThermPot}) and~(\ref{dielfunc}). However, they dominate transport and density inhomogeneity at very low densities where $k_F|z| \lesssim 1$ and hence the long-range nature of the Coulomb disorder becomes important at low densities near the Dirac point.

The additional screening effect in DLG is the strongest for impurities located on top of the ``control'' layer for which the inhomogeneity varies by an order of magnitude, dashed line in Fig.~\ref{fig1:rmsns}(b). However, the screening is substantially less efficient for the impurities located in close vicinity of the ``studied'' layer and for the high concentration of impurities in SiO$_2$ substrate, see the solid line in Fig.~\ref{fig1:rmsns}(b).  

Note that our estimate of the density inhomogeneity relies on the assumption that locally electron density resembles quasiparticles in clean graphene. This is unjustified in the case of strong disorder for which $k_F\ell \sim 1$, see Sections~\ref{sec:transrate} and~\ref{sec:momentumrate}.

\section{Elastic relaxation rate and Boltzmann resistivity}

\subsection{Transport relaxation rate due to charged defects}\label{sec:transrate}

\begin{figure}
\includegraphics[width=0.4\textwidth]{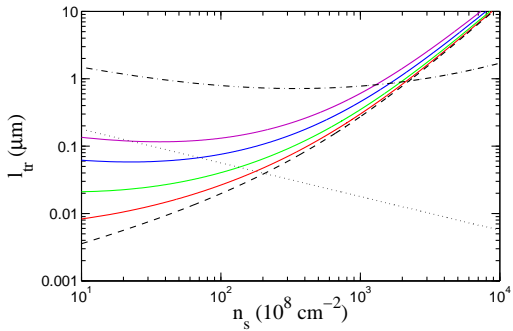}\caption{(color online) Transport scattering length in the ``studied'' layer. Solid lines from bottom to top (red, green, blue and magenta) correspond to $n_c=10^9, 10^{10}, 10^{11}, 10^{12}cm^{-2}$, with $z=-20nm$, $n_{i}=10^{12}cm^{-2}$. Dashed line corresponds to isolated monolayer graphene with the same parameters. For comparison the effect of a relatively small concentration of charged defects, $n_i=10^{11}cm^{-2}$, near ``studied'' layer, $z=-1nm$ is shown as dash-dotted line. The dotted line corresponds to $k_F\ell_{tr}=1$ where quantum and classical parts of the conductivity are of the same order. }\label{fig3:taup}
\end{figure}

Additional screening in DLG leads to a substantial suppression of the transport scattering rate which in Born approximation reads,
	\begin{equation}
 	\frac{1}{\tau_{tr}^C}=\frac{2\pi}{\hbar}n_{i}\sum_{\mathbf{p'}}\left(1-\cos\varphi\right)|\langle \mathbf{p'}|V|\mathbf{p}\rangle|^2
\delta\left(\epsilon_\mathbf{p}-\epsilon_\mathbf{p'}\right).~\label{eq:tau}
	\end{equation}
where $\varphi$ and $q\equiv|\mathbf{p-p'}|=2k_F|\sin\frac{\varphi}{2}|$ are the scattering angle and momentum of electrons on the Fermi surface in the ``studied'' layer, and $|\langle \mathbf{p'}|V|\mathbf{p}\rangle|^2=\frac{1}{2}(1+\cos\varphi) |\frac{\upsilon_q}{\varepsilon(q)}|^2e^{-2q|z|}$ is the matrix element of the Coulomb potential of a defect in graphene. Fig.~\ref{fig3:taup} shows the variation of the transport mean free path approximated by Eq.~(\ref{eq:tau}) within experimentally accessible densities in the ``control'' layer $10^9cm^{-2}\lesssim n_c \lesssim 10^{12}cm^{-2}$.

\subsection{Momentum relaxation rate: location of impurities}\label{sec:momentumrate}

\begin{figure}
\includegraphics[width=0.25\textwidth]{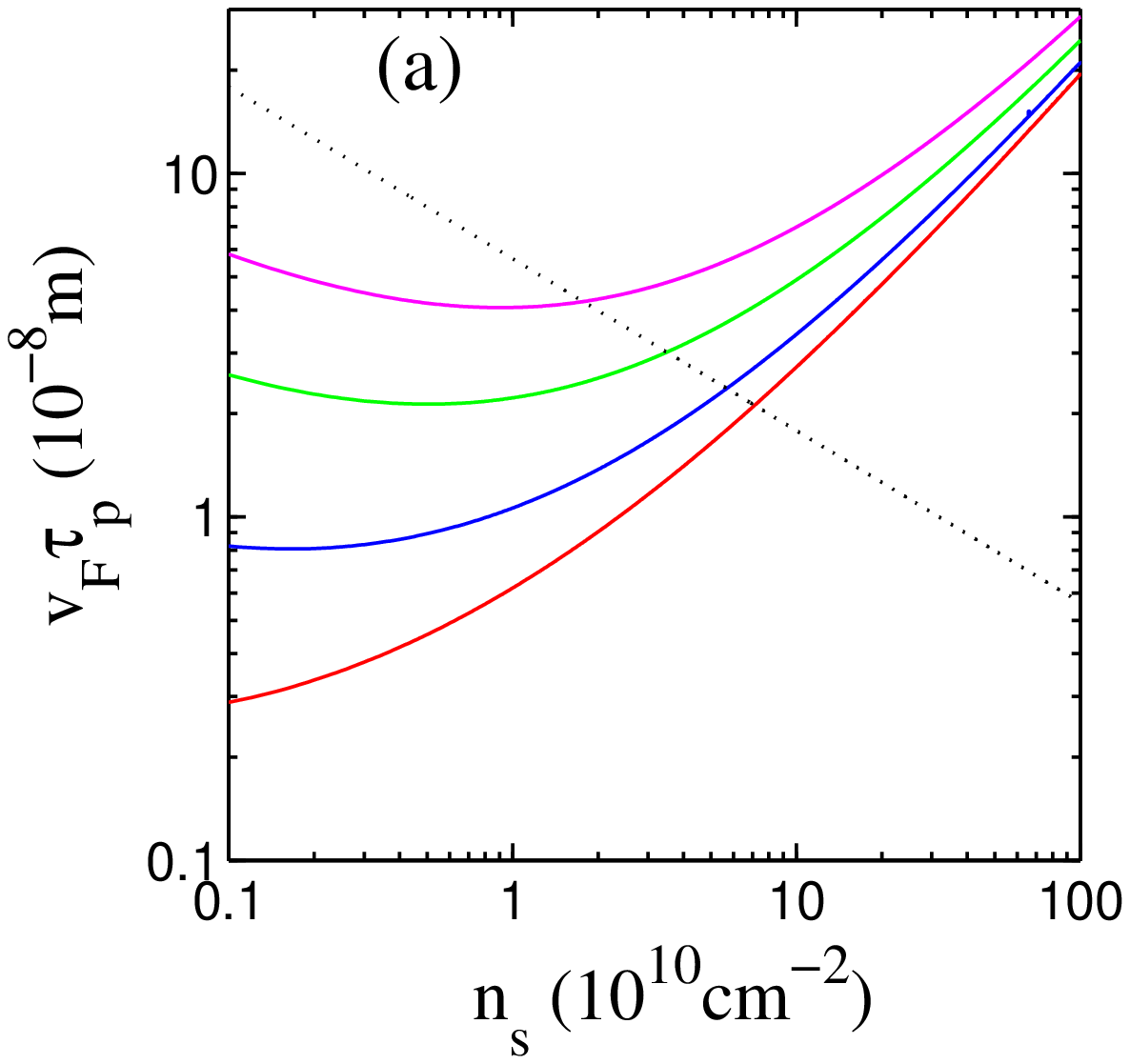}
\includegraphics[width=0.25\textwidth]{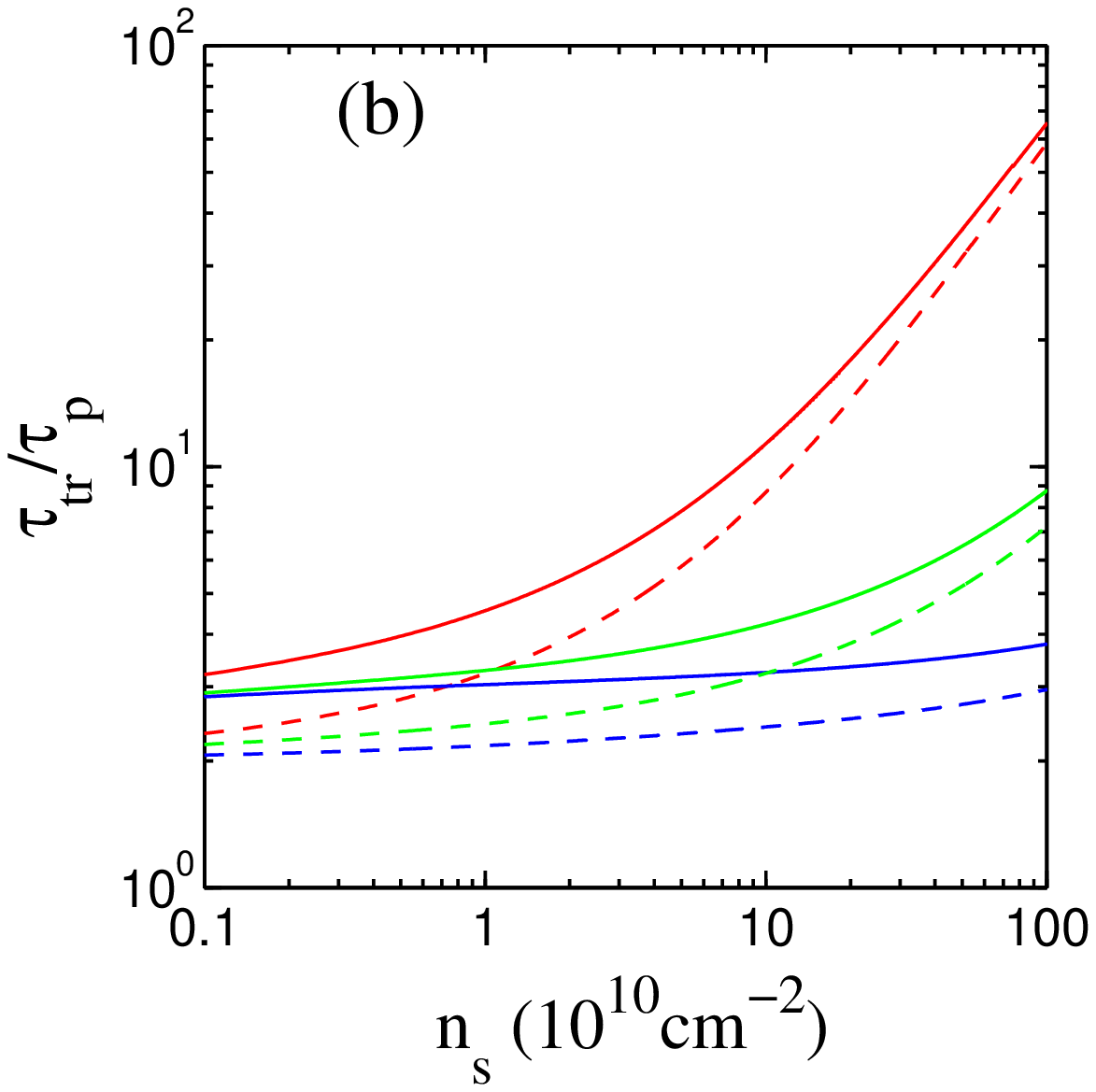}
\caption{ (color online) (a) Momentum relaxation rate as a function of the density in the ``studied'' layer $n_i=10^{12}cm^{-2}, z=-20nm$. Solid lines from bottom to top (red, blue, green and magenta) correspond to $n_c=10^9,10^{10},10^{11},10^{12}cm^{-2}$ respectively. Dotted line corresponds to $k_F \ell =1$ where Born approximation becomes insufficient. (b) Ratio of the transport and momentum relaxation times. Solid lines from bottom to top (blue, green and red) correspond to  ($n_i=10^{11}cm^{-2}, z=1nm$), ($n_i=10^{11}cm^{-2}, z=5nm$) and ($n_i=10^{12}cm^{-2}, z=-20nm$) respectively with $n_c=10^{12}cm^{-2}$. Dashed lines correspond to the same parameters with $n_c=10^{8}cm^{-2}$. }\label{fig3:rateratio}
\end{figure}

Different spatial configurations of impurities within DLG may be distinguished experimentally by analyzing the ratio of the momentum relaxation rate to the  transport scattering rate. These ratios are substantially different in the cases of scattering on long range potential of remote and nearby impurities. The transport rate, Eq.~(\ref{eq:tau}), is mostly determined by backscattering with large momentum transfer, and therefore decays quickly with the distance $z$ between the charged defects and the ``studied'' layer. The momentum relaxation rate, given by,
\begin{gather}
\frac{1}{\tau_p} = \frac{2\pi}{\hbar}n_{i}\sum_{\mathbf{p'}}|\langle \mathbf{p'}|V|\mathbf{p}\rangle|^2
\delta\left(\epsilon_\mathbf{p}-\epsilon_\mathbf{p'}\right),~\label{eq:taup}
\end{gather} 
includes forward scattering with small momentum transfer which is only weakly dependent on $z$, see Fig.~\ref{fig3:rateratio}(a). The ratio of the momentum relaxation rate to the transport rate is shown in Fig.~\ref{fig3:rateratio}(b) for the three different location of impurities considered in Sec.~\ref{sec:locImp}. Note that Born approximation used in Eq.~(\ref{eq:taup}) is insufficient when $k_F\ell_p \sim 1$, $\ell_p\equiv v_F\tau_p$ shown as dotted line in Fig.~\ref{eq:taup}(a) and further corrections to the elastic relaxation rate have to be included. In the case of remote impurities in SiO$_2$ there is a peculiar regime where $k_F\ell_p \sim 1$ and the Born approximation is insufficient yet $k_F\ell_{tr}\gg 1$ which suggests diffusive transport. Note however that the perturbative analysis presented here is quantitatively correct only when $k_F\ell_p \gg 1$.

\subsection{Boltzmann resistivity}

\begin{figure}
\includegraphics[width=0.225\textwidth]{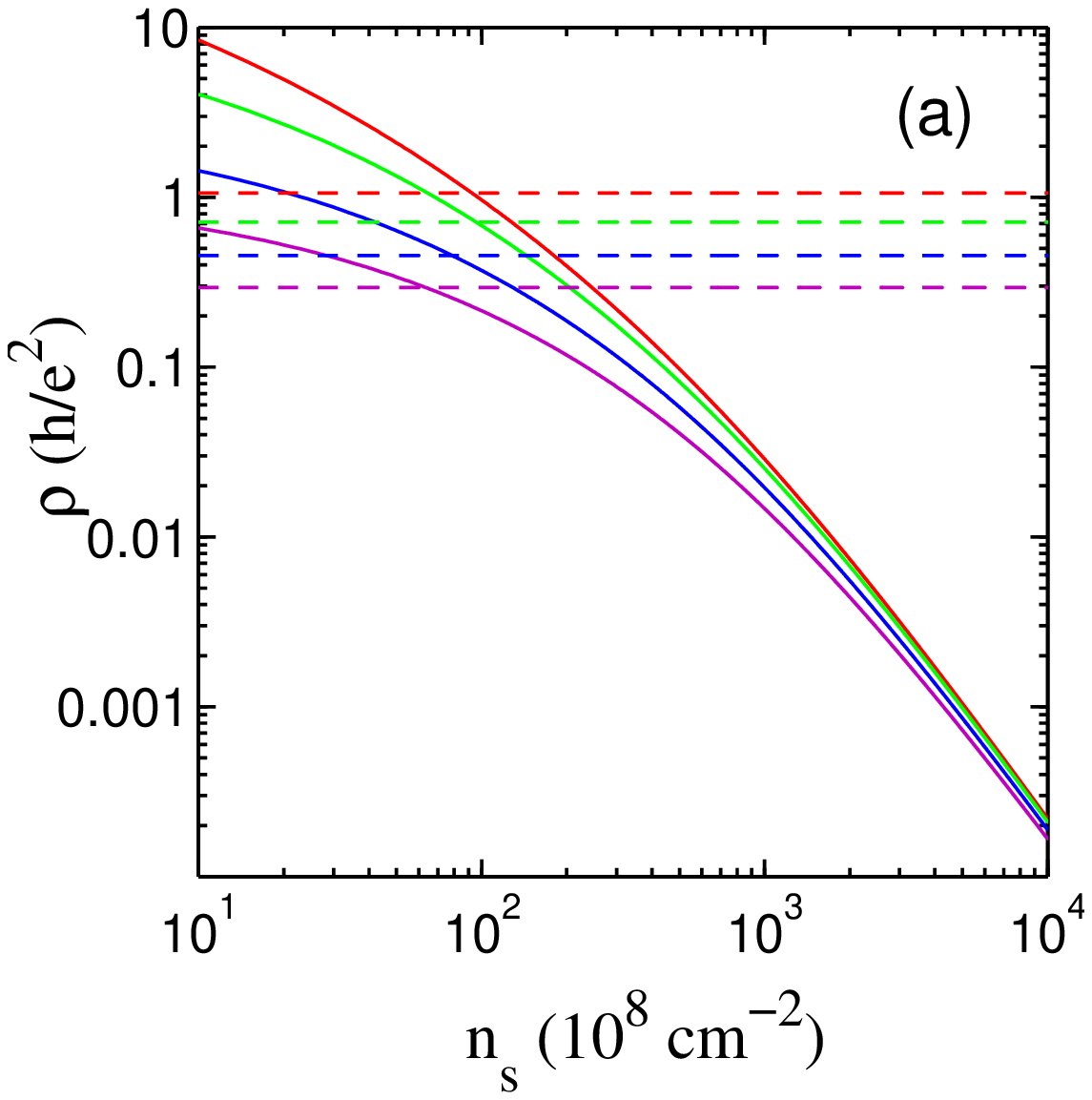}
\includegraphics[width=0.225\textwidth]{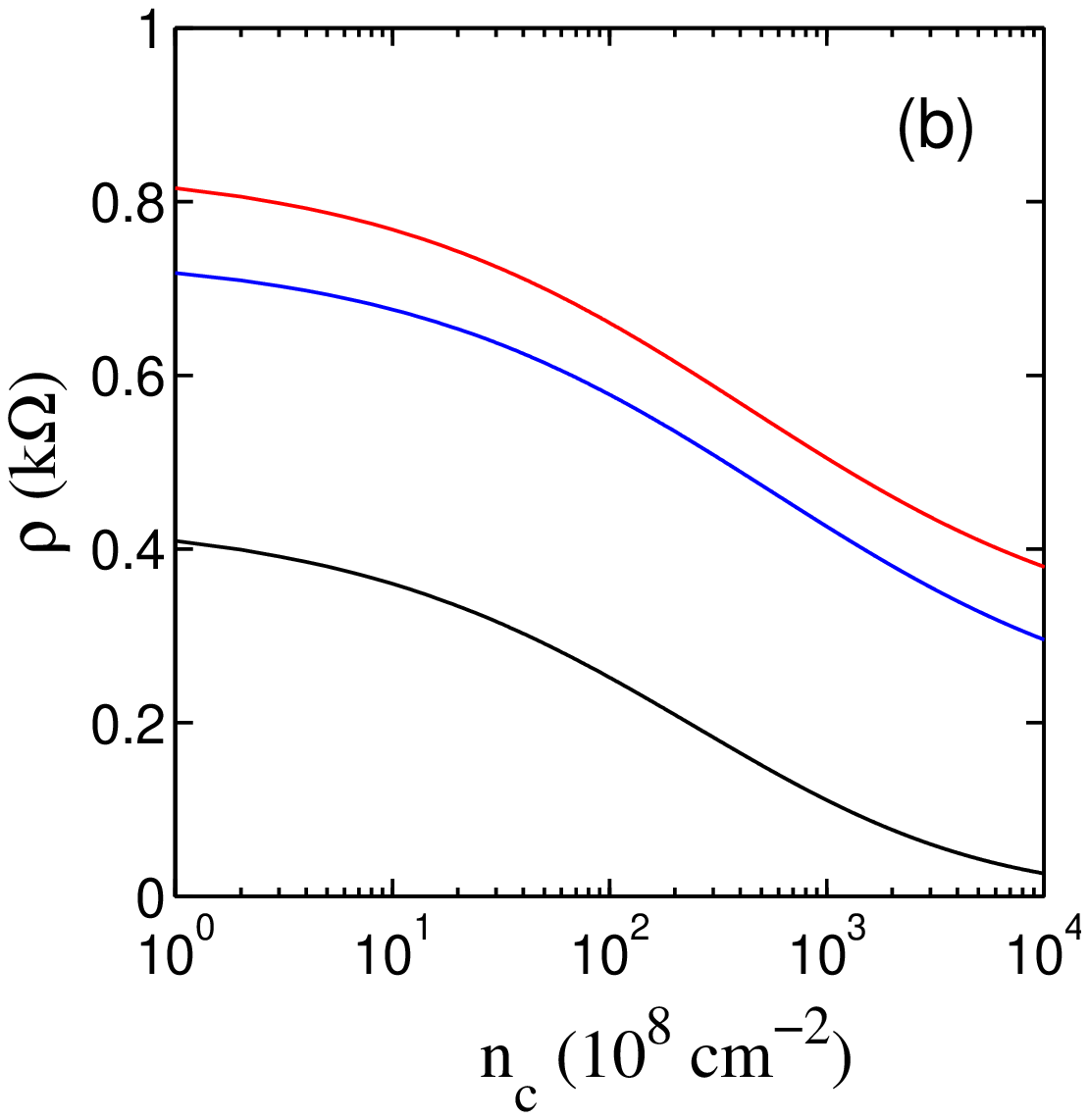}
\caption{ (color online) Resistivity of the ``studied'' layer $\rho(n_{s},n_{c})$. (a) Solid lines from top to bottom (red, green, blue and magenta)  correspond to $n_{c}=10^{9},10^{10},10^{11},10^{12}cm^{-2}$ respectively, with  $n_{imp}=10^{12}cm^{-2}$, $z=-20nm$, $d=4nm$. Dashed lines correspond to inhomogeneity induced maximum resistivity with the same parameters from top to bottom. (b) Resistivity at intermediate density $n_s=10^{11}cm^{-2}$. Top (red) line corresponds to $n_{imp}=10^{12}cm^{-2}$, $z=-20nm$, bottom (black) and middle (blue) lines correspond to $n_{imp}=10^{11}cm^{-2}$ and $z=5nm$, $1nm,$ respectively.} \label{fig4:resistivity}
\end{figure}

Suppression of the relaxation rate is reflected in the Boltzmann resistivity $\rho(n_s,n_c) = \left(\frac{2e^2}{h} k_F\ell_{tr}\right)^{-1}$ shown in Fig.~\ref{fig4:resistivity}(a). It is the strongest near the neutrality point in the ``studied'' layer in which case both, $n_s \ll n_c$ and $n_s \ll \frac{1}{\pi z^2}$ so that the high concentration of defects in SiO$_2$ substrate dominates the transport properties. In the formal limit $2k_F^{(1)} =2\sqrt{\pi n_s} \ll \kappa_2=8\pi e^2 \nu_c$, ($\kappa_{2}$ is the Thomas-Fermi screening wavevector in the ``control'' layer) the charged impurity potential becomes effectively short ranged which leads to the resistivity independent of the density $n_s$,
	\begin{equation}
	\rho_{max}(n_c) =\frac{\pi}{16} \left(\frac{h}{2e^2}\right)\frac{n_{i}}{n_c}\left(1+4\sqrt{\pi n_c}\alpha x\right)^{2},\label{rhomax}
	\end{equation}
where $x=\left(|z-d|-|z|+d\right)$. This strong-screening limiting resistivity, however, is unlikely to be reached, since within the metallic regime and accessible densities $k_F^{(1)}/\kappa_{2}$ is of the order of unity. Moreover, the value of the resistivity at low densities $n_s$ is affected by the doping inhomogeneity. The resistivity maximum in the presence of density inhomogeneity (electron-hole puddles) can be estimated~\cite{RRNFalko07,DasSarmaRMP11} using the effective medium theory~\cite{EMT1,*EMT2}, which allows averaging of the resistivity of a system split into regions with random values of resistance characterized by a distribution $P[\rho(n)]$. The averaged resistance, $\rho_{EMT}$, in this case is given by the solution of a self-consistent effective medium equation,
	\begin{gather}
	\int dn P[n]\frac{\rho(n)-\rho_{EMT}}{\rho(n)+\rho_{EMT}}=0.
	\end{gather}
We approximate the distribution of the density fluctuations by a Gaussian with zero average and variance given by $n_{rms}=\langle V^4\rangle$, which compares well with full Thomas-Fermi-Dirac approach~\cite{AdamRossi09} in the case of disordered monolayer graphene. The results are shown in Fig.~\ref{fig4:resistivity}(a) as dashed lines. Note however that due to the relatively weak doping inhomogeneity induced by the remote impurities in SiO$_2$ the value of the metallicity parameters $2\lesssim k_F(n_{rms})\ell_{tr} \lesssim 4$ and $k_F\ell_p \sim 1$ indicate that quantum corrections could be important in this low-density regime. However, the long-range character of the disorder potential makes quantitative analysis of quantum interference difficult~\cite{RossiBardarsonUCF,AleinerEfetov06,OstrovskyMirlin06} in this regime. 

At higher density $n_s$ the random inhomogeneity can be neglected and transport is metallic, therefore the linear screening model and the Boltzmann resistivity provide an accurate description of the system. At even higher densities $n_s \pi z^2\gtrsim 1$ the exponential factors in Eqs.~(\ref{dielfunc}, \ref{eq:tau}) lead to the strong suppression of the effect of remote impurities in SiO$_2$. As a result charged defects in the vicinity of graphene layer and neutral scatterers determine the resistivity. Finally, at the very high density $k_Fd\gtrsim1$ the typical monolayer behavior of the resistivity is restored, see Fig.~\ref{fig4:resistivity}(a). 

\subsection{Intermediate density regime}\label{sec:IntermedDensity}

There exists a regime of densities $n_s$ where Eqs.~(4,5) provide a quantitatively accurate description of the dependence of the elastic relaxation rate on both densities $n_s$ and $n_c$. Formally, the constraints defining this regime are: (i) the effect of the density inhomogeneity on the resistivity is negligible, $\sqrt{\overline{\left(n_s-\overline{n_s}\right)^2}} \ll \overline{n_s}$, where $\overline{O}$ stands for spatial average of $O$; (ii) transport in the ``studied'' layer is metallic $n_s \gg \frac{1}{\pi \ell^2_p}$. In addition we assume (iii) that the density in the ``studied'' layer is sufficiently small, $n_s \ll n_c $, so that the resistivity changes substantially with changing $n_c$.  Here $n_c$ is limited by the experimentally accessible density range $n_c\lesssim 5\times10^{12} cm^{-2}$; and (iv) the trigonal warping of the Fermi line has a negligible effect on weak localization corrections~\cite{Kechedzhi06}, $\tau_w^{-1}/\tau^{-1} =2\left(\pi\mu \hbar \tau n_s\right)^2 \ll 1$, with $\mu\equiv\frac{v_Fa}{4}$ and $a\approx 1.42$\AA is the distance between the nearest-neighbors in the hexagonal lattice of graphene. The latter constraint simplifies the analysis of characteristics of various scattering mechanisms that determine the quantum transport in the system. (v) Scattering due to Coulomb defects remains dominant $k_Fz \lesssim 1$. Throughout the text we refer to the range of densities $n_s$ where all constraints (i)-(v) are satisfied as the ``intermediate density regime''. 

The intermediate density regime can be used for a quantitative analysis of both Boltzmann and weak localization contributions to the resistivity. The effect of screening electron density $n_c$ in the ``control'' layer on the resistivity is illustrated in Fig.~\ref{fig4:resistivity}(b). The effect of $n_c$ on the weak localization correction in the DLG in the intermediate density regime is discussed below. 

\section{Quantum correction}

We generalize the perturbative theory of weak localization in graphene~\cite{Kechedzhi06,EPJKechedzhi} to the DLG heterostructure in the intermediate density regime defined in Sec.~\ref{sec:IntermedDensity} above.

\subsection{Perturbative theory of weak localization}

We briefly review the theory of weak localization correction to the resistivity of graphene. Weak localization originates from the interference contribution to the probability to scatter backwards. The latter is given by the modulus square of the sum of the quantum mechanical amplitudes $A_i$ associated with each possible backscattered trajectory, $P\sim|\sum A_i|^2$. This contains the classical part, $P_{cl}\sim\sum|A_i|^2$, and the quantum interference contribution, $\delta P \sim \sum A_iA_j^{\ast}$. The latter vanishes for generic trajectories. However, in case $A_j^{\ast}$ is a time reversal image of $A_i$, which is possible for loop-shaped self-intersecting trajectories, the phase factors cancel exactly, $\delta P \sim \sum'|A_i|^2$. Therefore the average contribution of this special type of trajectories to the backscattering probability is non-zero. $\delta P$ for a given self-intersecting trajectory is proportional to the ratio of the width of the trajectory $v_F\lambda dt $ to the 2D volume $Dt$ that can be spanned by diffusion during time $t$. The resulting correction to the conductivity $\delta \sigma \sim \int \lambda v_F dt/(Dt) \sim \ln \tau_{\varphi}/\tau$ is proportional to the log of the ratio of the travel times along the longest $\tau_{\varphi}$ and the shortest $\tau$ trajectories and is typically of the order of the conductance quantum. The longest trajectories $\tau_{\varphi}$ that limit the extent of quantum interference are given by either the system size or by phase breaking effects. 

Two degenerate flavors (two valleys) of chiral Dirac fermions in graphene are characterized by an isospin degree of freedom that has a fixed projection on the direction of momentum. Therefore backscattering of chiral charge carriers is accompanied by the reversal of isospin which gives rise to an additional phase difference $\pi N$ between the two images of a looped backscattered trajectory, where $N$ is the winding number of the trajectory. The interference contribution to conductivity in this case acquires an additional minus sign which results in anti-localization~\cite{Ando02,*Khvesh06,*MorpugoGuinea} effect in contrast to the usual weak localization in 2D~\cite{Bergmann1984}. In realistic devices the vector potential (intravalley) disorder, characterized by a phenomenological rate $\tau^{-1}_{z}$, breaks the symmetry between the two sublattices of graphene and therefore suppresses the anti-localization effect. Atomic scale disorder, characterized by $\tau^{-1}_{iv}$, scatters electrons between different valleys and therefore restores the localization sign of the interference correction. Quantitatively this effect is described by a perturbation expansion in $k_F\ell \gg 1$ which results in the conductivity correction~\cite{Kechedzhi06},
	\begin{equation}
	\delta \sigma = -\frac{e^2}{\pi h} \left[ \ln\left(1+2\tfrac{\tau_{\varphi}}{\tau_{iv}}\right)
	-2 \ln \frac{\frac{\tau_{\varphi}}{\tau_{tr}}}{1+\frac{\tau_{\varphi}}{\tau_{iv}}+\frac{\tau_{\varphi}}{\tau_{z}}} \right].\label{eq:wl}
	\end{equation} 
Here the first term on the left hand side describes intervalley interference sensitive only to the ratio $\tfrac{\tau_{\varphi}}{\tau_{iv}}$, the second term describes the intravalley interference which is also sensitive to smooth intravalley scattering rate $\tau_z^{-1}$ due to the vector potential disorder and the total transport scattering rate $\frac{1}{\tau_{tr}}\approx\frac{1}{\tau_{tr}^C}$. The left hand side in the latter equation, given by Eq.~(\ref{eq:tau}), is obtained modeling charged defects by a Gaussian disorder with a finite range in space and zero average of the potential. 

In Fig.~\ref{fig7:WLNS}(a) we show the lines in the parameter space that separate the negative $\delta\sigma <0$, localization-like correction from the positive one $\delta \sigma >0$. In the presence of strong vector potential disorder, $\tau_z \sim \tau_{tr}$ the intravalley interference term is strongly suppressed. This results in the weak localization sign of the correction even in the case of relatively weak intervalley scattering, see the blue line in Fig~\ref{fig7:WLNS}(a). By contrast, in the presence of relatively weak vector potential disorder $\tau_z^{-1}\ll\tau_{tr}^{-1}$, a substantially stronger intervalley scattering is required to compensate for the intravalley term in (\ref{eq:wl}) and for the correction to have the localization sign~\footnote{Corrugations of the graphene sheet give rise to an effective vector potential disorder and therefore increase $\tau_{z}^{-1}$. This effect is expected to be small in devices used in Ref.~\onlinecite{Geim11} since surface roughness of hBN substrate is an order of magnitude smaller than of SiO$_2$.}. We emphasize that these analytical considerations are entirely restricted to the perturbative regime, $k_F\ell_p \gg1$ of the quantum interference correction to the conductivity and do not apply to the strong localization regime of metal-insulator transitions.

\subsection{Intervalley scattering}

\begin{figure}
\includegraphics[width=0.23\textwidth]{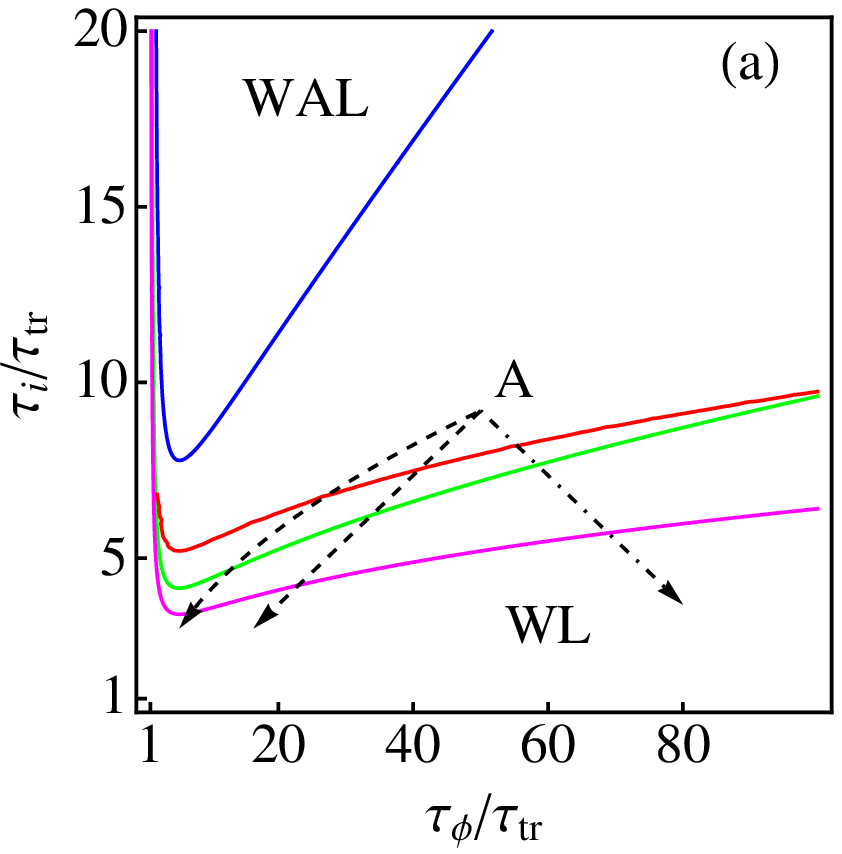}
\includegraphics[width=0.23\textwidth]{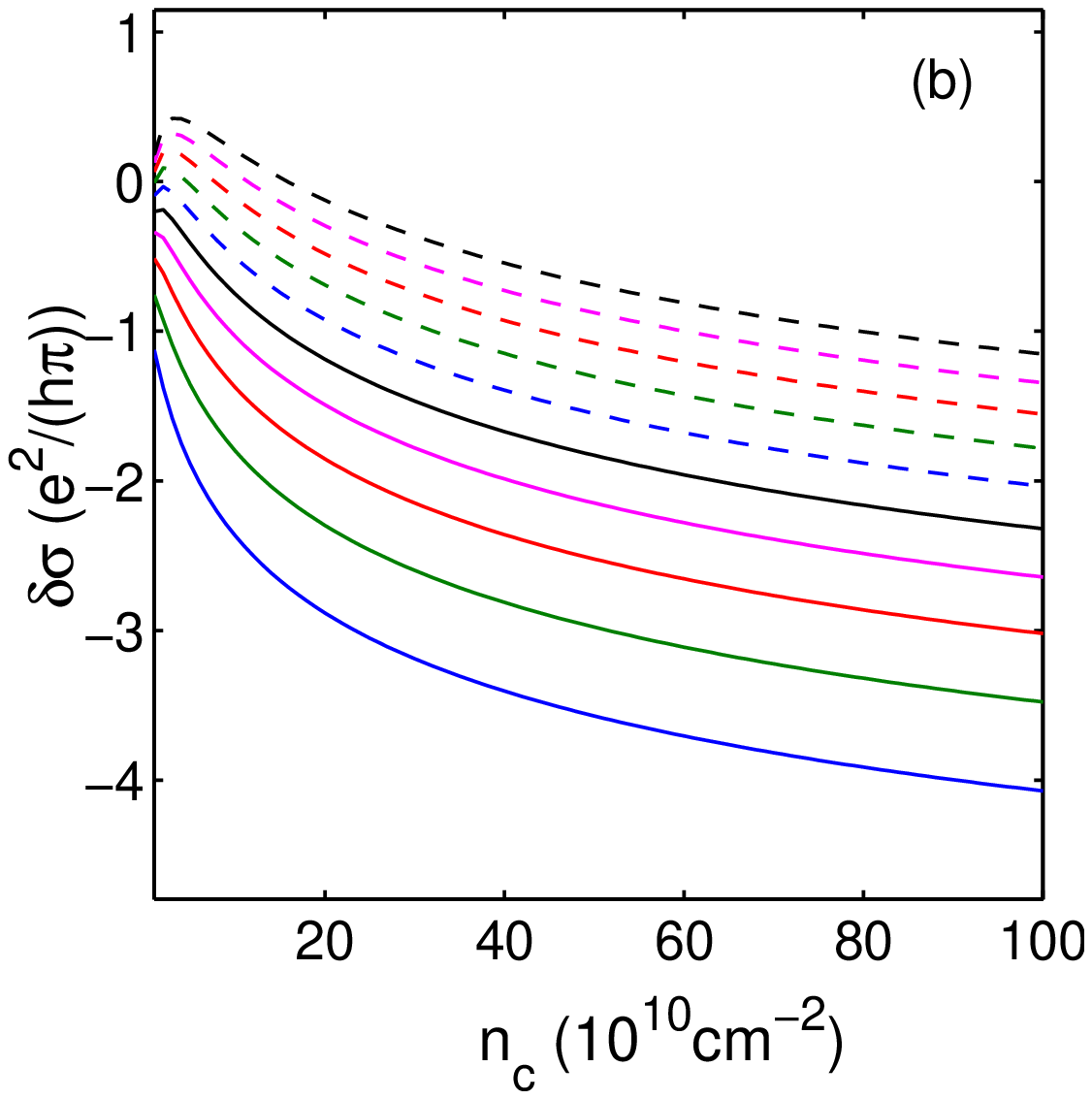}
\caption{ (color online) (a) Weak antilocalization to locaization crossover boundary in the parameter space of disorder in DLG. Magenta, green and blue lines correspond to the fixed values of $\tau_{\varphi}/\tau_z=100,10,3$ respectively. Red line corresponds to monovalent adsorbate disorder. The dashed lines show example trajectories that a DLG initially at point A takes with changing $n_c$ in case of fixed $\tau_{\varphi}$, straight line, and coherence limited by the system size, the curved trajectory $\sim \sqrt{\tau_{\varphi}/\tau_{tr}}$. Dash dotted line corresponds to the trajectory with coherence limited by electron-electron interactions in the diffusive regime Eq.~(\ref{dephasing}). (b) Weak localization correction for different values $ 0.1\mu m\leq v_F \tau_i\leq 1\mu m$ from bottom to top with $n_s=5\times10^{10}cm^{-2}$. We assume $n_i=10^{12}cm^{-2}$ at $z=-20nm$ and dephasing is limited by the electron-electron interactions in the diffusive regime $T\tau \ll 1$, Eq~(\ref{dephasing}). We took $v_F\tau_\varphi =0.3\mu m$ at $n_c =10^{10}cm^{-2}$.} \label{fig7:WLNS}
\end{figure}

Microscopic origin of the intervalley scattering at intermediate densities therefore deserves a more careful consideration. Electrostatic potential can give rise to a strong intervalley scattering only if its amplitude is larger than the bandwidth in which case the lowest order Born approximation is insufficient, see Appendix and Ref.~\onlinecite{BernevigBKTinMLG} for details. Such strong potentials can be induced by lattice defects and adsorbate molecules chemically bound to carbon atoms of the graphene lattice. Scattering characteristics of such defects depend strongly on the detailed nature of the electronic structure, type of the chemical bonding and the position of the defect in the unit cell of graphene~\cite{WehlingKatsnelson}. Such details are beyond the scope of the present analysis. Instead we employ a more generic phenomenological model of two types of scatterers namely resonant scatterers and short range disorder. The former describes the effect of monovalent chemical groups and vacancies in the graphene lattice~\cite{Basko08,RobinsonSchomerusFalko,TitovResScatt,*OstrovskyTitovResScatt}. The scattering rate in this case decreases with density $\tau_R^{-1}\sim \left(\sqrt{n}\ln n\right)^{-1} $ which in monolayer graphene is practically indistinguishable from that due to charged defects $1/\tau_{tr}^C \sim 1/\sqrt{n}$, see Ref.~\onlinecite{DasSarmaRMP11}. Short range scatterer model describes atomic defects that do not give rise to resonant scattering. The corresponding scattering amplitude grows with density $\tau^{-1}\sim \sqrt{n}$ and the resulting contribution to the conductivity is density independent. In the case of both sources of intervalley scattering a relatively weak $\tau_{iv}^{-1}$ is sufficient for the quantum correction to have the weak localization sign, Fig~\ref{fig7:WLNS}(a).

There exists substantial experimental evidence for the presence of intervalley scattering in graphene. In particular, the observation of negative weak localization magnetoresistance suggests a rather high intervalley scattering rate corresponding to $L_{iv}\equiv\sqrt{D\tau_{iv}}\sim0.1-0.3\mu m$ at high densities $n>10^{11}cm^{-2}$,~\cite{NadyaMason,Savchenko08,KoreanGroup,SPMImpurity,SavchenkoWLWAL,WLvEEI,EEIEpitaxial,BrunSchweigEpitax,deHeer,Korean2010,GateTunableValley}. Alternative measurements of a weak D-peak in the Raman spectrum demonstrated the presence of intervalley scattering at high energies $\sim 1$eV~\cite{DPeak}. More detailed quantitative characteristics of the intervalley scattering such as density dependence are not currently available experimentally. 

\subsection{Phase breaking effect}

The size of the self-intersecting trajectories that give rise to the interference correction is limited by either the system size or the phase breaking effect. The latter can be caused by scattering on magnetic defects~\cite{SiCWLMRFalko} or inelastic electron-phonon or electron-electron collisions~\cite{SavNarozhnyEE}. Both latter effects result in temperature dependent dephasing rate $\tau_{\varphi}^{-1}$, which is linear $\tau_{\varphi}^{-1} \sim T$ in the case of electron-phonon scattering and linear or quadratic in the case of electron-electron interactions in the diffusive and ballistic regime respectively. Experiments report typical values for the dephasing length $\sqrt{D\tau_{\varphi}}\sim 2 \mu m$ at the lowest temperatures $T\leq 1K$ \cite{NadyaMason,Savchenko08,KoreanGroup,SPMImpurity,SavchenkoWLWAL} and $L_{\varphi}\sim 0.2 \mu m$ at higher temperatures $T\sim20K$. More recent measurements on large area epitaxial graphene devices accessed very high temperature regime $T\sim 100K$ corresponding to a short dephasing length $L_{\varphi}\sim0.04\mu m$~\cite{SiCWLMRFalko,MirlinQFMLG}. The available experimental information about the density, disorder,  and temperature dependent phase breaking length in graphene is, however, not extensive, and more work is needed before definite quantitative statements can be made in details.

\subsection{Effect of screening on the dephasing rate due to inelastic electron-electron collisions}

In this section we estimate the effect of additional screening on the dephasing rate caused by inelastic electron-electron collisions with low energy transfer $\hbar\omega\lesssim k_BT$. We estimate the phase breaking rate up to a numerical coefficient using the self-consistent approach~\cite{WavesInRandommedia,NarozhnyAleiner02,AAbook,AkkermansBook,RelaxationRatesMirlin} which is sufficient to analyze the dependence of the phase breaking rate on the densities in the two layers of DLG. There are two distinct dephasing temperature regimes, diffusive, $k_B T \gg \hbar \tau_{tr}^{-1}$ , and ballistic, $k_BT \ll \hbar \tau_{tr}^{-1}$ which correspond to distinct temperature dependences of $\tau_{\varphi}^{-1}$. 

In the diffusive regime, $k_BT\tau_{tr} \ll \hbar$, electrons undergo diffusive motion which increases the time they spend in vicinity of each other (where interaction is the strongest) and therefore enhances the inelastic scattering rate. Electron-electron collision rate is given by Fermi golden rule,
\begin{gather}
\frac{1}{\tau_{in}} = 4\pi \nu^3 \int_{-\infty}^{\infty} d\omega d\epsilon' F(\epsilon,\epsilon',\omega) W^2(\omega), \label{tauIn} \\
F(\epsilon,\epsilon',\omega)=f_\epsilon'(1-f_{\epsilon-\omega})(1-f_{\epsilon'+\omega})+(1-f_\epsilon')f_{\epsilon-\omega}f_{\epsilon'+\omega}, \nonumber
\end{gather}
where $f_\epsilon \equiv 1/(1+\exp(\beta\epsilon))$ and $\beta^{-1}=k_BT$. The interaction kernel in~(\ref{tauIn}) is given by,
\begin{gather}
W^2(\omega)=\frac{1}{\nu^4}\sum |\langle \alpha\gamma|U_{11}|\beta\delta\rangle|^2 \\
\times
\delta(\epsilon-\epsilon_\alpha)\delta(\epsilon'-\epsilon_\gamma)
\delta(\epsilon-\omega-\epsilon_\beta)\delta(\epsilon'+\omega-\epsilon_\delta). \nonumber
\end{gather}
The matrix element of the screened interaction is written in terms of exact eigenstates $|\alpha\rangle$ of non-interacting Dirac Hamiltonian in the presence of a given disorder potential. The screened electron-electron interaction potential in DLG is given by $U_{11}(q,\omega)=\frac{\upsilon_q}{\varepsilon_0(q,\omega)}$ which includes the dynamical dielectric function of Eq.~(\ref{dielfunc}) where we need to put $z=0$ for interaction within the ``studied'' layer. 

The expression for the collision rate~(\ref{tauIn}) has to be averaged over disorder realizations. The result of this averaging can be then expressed in terms of the diffusion propagator in graphene $D(q,\omega)$ which describes the transport of the charge density through the disordered system. The diffusion propagator is found by solving the diffusion equation $\left(\partial_t - \Delta D_1\right)D(\mathbf{r},\mathbf{r'}) = \delta(\mathbf{r-r'})$ in the ``studied'' layer~\cite{KechedzhQKE}. Here $D_i$ is the diffusion coefficient in the ``studied'' ($i=1$) or ``control'' ($i=2$) layer respectively. As a result the Eq.~(\ref{tauIn}) at $q\ell_1\ll1$ and $\omega\tau_{tr}\ll 1$ takes the form,
\begin{gather}
\frac{1}{\tau_\varphi} = \frac{8\nu}{\pi} \int_{\tau_\varphi^{-1}}\frac{d\omega d^2q}{(2\pi)^2} \frac{\omega|U(q,\omega)|^2}{\sinh\omega}\left[\Re D(q,\omega)\right]^2\label{DiffPhi}
\end{gather} 
where we keep only the logarithmically divergent terms. Also we introduce a low energy cutoff $\omega\sim\tau_{\varphi}^{-1}$ which reflects the fact that only collisions with large enough energy transfer $\omega\gtrsim \tau_{\varphi}^{-1}$ produce a substantial dephasing of the electrons with diffusion time limited by $\tau_{\varphi}$~\cite{OmegaGrtrTauphi,AAbook,AkkermansBook}. Solving~(\ref{DiffPhi}) self-consistently gives an estimate for the dephasing rate.

At high densities such that ${q\ell_i \gg 1}$, $i=1,2$, the screening is ballistic for which case we can approximate $q\upsilon_q\chi_{ii}(q,0)\approx \kappa_i$ and Eq.~(\ref{DiffPhi}) gives,
\begin{gather}
\frac{1}{\tau_\varphi}=\tilde{C}\frac{T}{8\pi\nu_1 D_1}\ln \pi \nu_1 D_1,\label{dephasing}
\end{gather}  
with
\begin{gather}
\tilde{C}\approx
	\frac{1}{2}\frac{\zeta^{2}}{\left(\zeta+\frac{\kappa_2}{\kappa_{1)}}\right)^{2}}. 
	\label{Cball}
\end{gather} 
where we introduce the screening parameter $\kappa_i=8\pi e^2\nu_i$ for $i$-th layer and $\zeta\equiv1+2d\kappa_2$.

At lower densities the polarization operator acquires a diffusion pole,
\begin{gather}
\chi_i(q,\omega) \approx g \nu_i \left(1+i\omega\frac{1}{-i\omega+D_iq^2}\right). 
\end{gather}
The presence of the diffusion pole affects only the coefficient in Eq.~(\ref{dephasing}).
In the case of short screening length $\kappa_i=8\pi e^2\nu_i > \ell_{tr}^{-1}$ we obtain for  
the coefficient $C\rightarrow\tilde{C}$, 
\begin{gather}
C=\frac{1}{1+\frac{\kappa_2}{\kappa_1}\frac{D_{2}}{D_{1}}}\frac{\left(\zeta+\frac{\kappa_2}{\kappa_1}\right)+\frac{D_{2}}{D_{1}}\zeta^{2}\left(1+\frac{\kappa_2}{\kappa_1}\frac{D_{2}}{D_{1}}\right)}{\left(\zeta+\frac{\kappa_2}{\kappa_1}\right)\left(1+\left(\zeta+2\frac{\kappa_2}{\kappa_1}\right)\frac{D_{2}}{D_{1}}\right)}. \label{diffusivePhi}
\end{gather}
In the limit of $\kappa_2/\kappa_1\gg 1$ the results (\ref{diffusivePhi}) and (\ref{Cball}) both approach,
\begin{gather}
C \approx \frac{1}{2} \frac{\kappa_1^2}{\kappa_2^2}.
\end{gather}
The dephasing rate~(\ref{dephasing}) is strongly dependent on the screening density $n_c$ in the ``control'' layer through the diffusion coefficient $D_1$ and the factor $C$ or $\tilde{C}$.

In the high temperature $T\tau_{tr} \gg 1$ ballistic regime the polarization operator is approximated by the free particle form~\cite{Guinea06,HwangDasSarma2007}, $\chi_i(q,\omega)\approx g\nu_i \left(1+i\frac{\omega}{v_Fq}\right)$ at $\omega/(v_Fq)\ll1$. The dephasing rate in this case is given by~\footnote{The authors are grateful to Quizi Li for sharing unpublished results.},
\begin{gather}
\frac{1}{\tau_{\varphi}} \approx
\frac{\left(k_B T\right)^2}{16 \nu_1 \left(\zeta + \frac{\kappa_2}{\kappa_1}\right)}\ln\frac{k_BT}{16\pi v_F^2\nu_1},\label{ballisticPhi}
\end{gather}
where parameters $\kappa_i=8\pi e^2\nu_i$ and $\zeta\equiv1+2d\kappa_2$ have the same meaning as above and we assume $\kappa_id\ll1$.

\subsection{Gate tunable quantum interference correction}

In the presence of mixed scattering sources the suppression of the transport scattering rate with growing $n_c$ results in the increased relative strength of neutral scattering mechanisms. This is a simple consequence of the Coulomb scattering being effectively screened out by the ``control'' layer. As a result the amplitude of the quantum interference correction varies as a function of $n_c$. To illustrate this we assume a large concentration of charged defects $n_i$ located in the SiO$_2$ substrate and a small concentration of resonant scatterers $\tilde{n}_i$. Then, we have $\tau_{tr}^{-1}\approx\tau_C^{-1}+2\tau^{-1}_{iv}$ where we use $\tau^{-1}_{z}\approx\tau_{iv}^{-1}$ valid for a small concentration of single site defects.  In the present analysis we ignore the effect of screening by the ``control'' layer on the shift of the chemical potential in graphene due to charged defects.

Depending on the mechanism limiting the phase coherence a variety of behaviors of the conductivity correction as a function of the screening density $n_c$ is possible, see Fig.~\ref{fig7:WLNS}(a). For screening independent $\tau_{\varphi}^{-1}$ due to dephasing by magnetic impurities the system initially at point "A" in Fig.~\ref{fig7:WLNS}(a) takes a trajectory represented by the straight dashed line. In smaller devices coherence is more likely to be limited by the system size, in which case the effective rate $\tau_\varphi^{-1}\equiv\frac{D}{L^2}=\frac{v_F\ell_{tr}}{2L^2}$, depends on the diffusion coefficient and hence is sensitive to the screening effect of the ``control'' layer. This regime corresponds to the curved dashed trajectory in Fig.~\ref{fig7:WLNS}(a). The effect of screening by the ``control'' layer is the strongest in the case of dephasing limited by electron-electron interactions. In this case the ratio $\tau_{\varphi}/\tau_{tr}$ depends on $n_c$ through the coefficient $C$ in Eq.~(\ref{dephasing}), and the system takes the trajectory shown as the dash-dotted line in Fig.~\ref{fig7:WLNS}(a). This is the situation most likely to be realized in experiments on large exfoliated flakes of graphene. Dependence of the weak localization correction on $n_c$ for this case is given by Eq.~(\ref{eq:wl}) combined with (\ref{eq:tau}) and (\ref{diffusivePhi}), which is shown in Fig.~\ref{fig7:WLNS}(b). The effect of additional screening by ``control'' layer therefore expands the parameter range in which weak-localization sign of the quantum correction could be observed, and thus the possibility of a transition from the weak anti-localization to weak localization as a function of the ``control'' layer density (i.e. gate-tunable) in the nominally metallic studied layer is very high in this case.  This predicted transition has not yet been seen in experiments, but may actually exist in the samples of Ref.~\onlinecite{Geim11} in the metallic regime of the studied layer.

Weak field magnetoresistance allows an unambiguous measurement of the weak-localization correction. At small fields $B \ll B_{\varphi} = h/e(D\tau_{\varphi})$ the magnetoresistance can be approximated by,
	\begin{gather}
	\delta \sigma  \approx \frac{e^2}{24\pi h} \left( \frac{B}{B_{\varphi}}\right)^2
	\Phi\left(\frac{\tau_\varphi}{\tau_{iv}},\frac{\tau_\varphi}{\tau_z}\right), \\
	\Phi\left(\frac{\tau_\varphi}{\tau_{iv}},\frac{\tau_\varphi}{\tau_{z}}\right)\equiv \left[ 1 - \frac{1}{\left(1+2\frac{\tau_{\varphi}}{\tau_{iv}}\right)^2} - \frac{2}{\left(1+\frac{\tau_{\varphi}}{\tau_{iv}}+\frac{\tau_{\varphi}}{\tau_z}\right)^2} \right].\nonumber
	\end{gather}
In this regime we expect a gate tunable crossover from weak anti-localization to weak localization magnetoresistance to be observable especially at higher temperatures.  

\section{Comparison with experiment}

In this section we discuss the relation of the theory presented above to the recent transport measurements in DLG~\cite{Geim11}. The clever DLG design of Ref.~\onlinecite{Geim11} provides an opportunity to use the additional gate tunability of the resistivity in the ``studied'' layer, 
\begin{gather}
\rho(n_s, n_c)=\rho_C(n_s, n_c)+\rho_R(n_c),
\end{gather}
to distinguish the contributions of charged $\rho_C(n_s, n_c)$ and neutral $\rho_R(n_c)$ scattering mechanisms in the intermediate density regime, see Sec.~\ref{sec:IntermedDensity}. The latter can be deduced from the dependence of the mobility of the ``studied'' layer, $\mu(n_c)$, on the ``control'' layer density reported in the Supplementary Material of Ref.~\onlinecite{Geim11}. On the one hand these data are consistent with the presence of a high concentration of charged defects $n_i \approx 10^{12}cm^{-2}$ at the surface of SiO$_2$ substrate located at $z=-20nm$ away from the ``studied'' layer~\cite{Geim11}. On the other hand a similar behavior could result from a smaller concentration of charged defects $n_i\approx 10^{11}cm^{-2}$ in close vicinity of the ``studied'' layer or on top of the ``control'' layer, see Fig.~\ref{fig4:resistivity}(b) for comparison. The two situations may be distinguished experimentally by comparing the transport and momentum relaxation rates, see Fig.~\ref{fig3:rateratio}(b).  At low carrier densities however the precise location of impurities does not play an important role and therefore the substantially larger concentration of defects likely to be present in SiO$_2$ dominates the transport properties. 

The situation is different in the case of density inhomogeneity. We find that DLG is relatively inefficient in screening out electron-hole puddles induced by the high concentration of defects in SiO$_2$ as well as defects in close proximity to the ``studied'' graphene layer itself. Both allow some ``control'' in the experiments, the former can be reduced by changing the substrate and the latter may be reduced by improving the sample preparation. Moreover, both improvements were recently implemented in suspended devices~\cite{SuspGeim12}.

In the absence of more systematic data we can estimate the residual mobility associated with neutral scattering mechanisms from the maximum of the mobility curve, $\mu(n_c)$ at the maximum density in ``control'' layer reported in the Supplementary Material of Ref.~\onlinecite{Geim11}. The result is $\mu (n_s=10^{11}) \sim 1.2\times 10^5 cm^{-2}V^{-1}s^{-2}$, which corresponds to a relatively small density of resonant scatterers, such that $v_F\tau_{R} \gtrsim 0.8\mu m$ suggesting that the DLGs analyzed in Ref.~\onlinecite{Geim11} may demonstrate gate tunable weak anti-localization to localization crossover at intermediate densities.

Finally, we discuss the effect of coupling between graphene and hBN substrate which gives rise to spectacular new features in the electron spectrum at high densities~\cite{hBNNewDiracCones}. However, at low carrier densities this effect is not expected to be dramatic. Transport characteristics of graphene could be affected in three possible ways: (i) triangular symmetry of hBN breaks the sublattice symmetry of graphene and therefore gives rise to an additional trigonal warping of the Fermi surface~\cite{Brink}; our estimate shows negligible effect on the weak localization correction at low densities $n_s\lesssim10^{11}cm^{-2}$; (ii) hBN induced Moire pattern may dominate the density fluctuation at the neutrality point. This effect however has not been clearly identified in the local probe data~\cite{hBNInhomogeneity} to this point and therefore we expect it to be smaller than current levels of inhomogeneity induced by charged defects. (iii) Additional intervalley scattering may be caused by strong coupling of graphene to hBN. However, existing experimental data and \textit{ab initio} calculations~\cite{HBNDFTKatsnelson} suggest a rather weak coupling which is not expected to give rise to strong intervalley scattering. Nevertheless, a more detailed analysis is required to completely rule out all of these possibilities.

\section{Discussion and Conclusion}

Our goal in this work has been a careful theoretical study of the screening by the ``control'' layer in determining transport properties of the studied layer in double-layer graphene systems experimentally introduced in Ref.~\onlinecite{Geim11}.  The problem of interest to us is both subtle and complex, involving many disparate aspects of electronic transport phenomena, and our study is at best an approximate one because of the highly complex nature of the problem.  In particular, we are interested in studying both semiclassical Boltzmann and quantum localization contributions to the DLG resistivity, including the ``control'' layer screening effect, on an equal footing as much as possible.  Since potential, intravalley and intervalley elastic scattering affect quantum localization properties qualitatively differently, our work must include all mechanisms in the presence of ``control'' layer screening with equivalent considerations for all scattering processes contributing to the semiclassical and the quantum part of graphene resistivity.  Since the phase breaking length arising from inelastic scattering processes is an important ingredient in determining localization effects arising from quantum interference, we consider electron-electron interaction induced inelastic phase decoherence effects in our theory, neglecting electron-phonon interaction which is known to be weak in graphene~\cite{PhononsHwangDasSarma}.  We also neglect all direct contributions to transport from inelastic scattering processes since our interest is relatively low-temperature transport where quantum interference may play a role rather than high-temperature transport where inelastic phonon scattering plays a role.  We investigate the role of the ``control'' layer screening on the density inhomogeneity (i.e. electron-hole puddles) near the Dirac point finding that screening by the ``control'' layer is ineffective in suppressing the electron-hole puddles arising from charged impurities in the SiO$_2$ substrate far from the ``studied'' graphene layer (in fact, this is an important qualitative new result of our work).  Finally, our work treats both long-range Coulomb disorder arising from random charged impurities in the environment and the short-range disorder arising from neutral atomic defects on an equal footing.  

Since the theory involves (at least) six different independent characteristic length scales (elastic mean free path due to long-range and short-range disorder and intra- valley and inter-valley scattering length, the inelastic phase breaking length, the Fermi wavelength, the density fluctuation correlation length, the thermal length, etc.), it is necessarily a complex problem necessitating various approximations focusing on different aspects of the transport phenomena.  We concentrate on low-temperature transport at relatively high "metallic" densities where the quantum interference contribution to the resistivity can be treated as a perturbative weak localization correction to the semiclassical Boltzmann resistivity.  This precludes us from commenting directly on the very low density nonperturbative insulating behavior (as observed in Ref.~\onlinecite{Geim11}) from the perspective of strong Anderson localization phenomenon, but our theory establishes the clear possibility of a  gate-tunable transition from weak anti-localization to weak localization behavior in graphene at metallic densities, which would be a necessary precursor to a possible strong localization crossover behavior at low densities.  Whether such a weak localization transition, which must precede any strong localization induced low density metal-insulator transition, is operational in Ref.~\onlinecite{Geim11} was unfortunately not studied there, and the prediction of such a gate-tunable transition in the weak localization behavior is an important new prediction of our work.  We emphasize that such transition must be studied at relatively low temperatures, and it may be missed at the higher temperature range (10-100K) of study used in Ref.~\onlinecite{Geim11}. In addition, the strong localization induced insulating behavior at low carrier density should be exponential in temperature, and not a power law as observed in Ref.~\onlinecite{Geim11}. Given our concrete theoretical predictions, we hope that future experiments on DLG systems will resolve the important question of a possible low-density quantum localization transition in clean graphene.  Our work shows that it is possible for the intervalley scattering in graphene to be strong enough to induce localization, but at the same time be weak enough not to strongly suppress high-density mobility, particularly in the presence of screening by the ``control'' layer.

In conclusion, we analyze the effect of gate tunable screening provided by the additional ``control'' layer in DLG on both classical and quantum parts of the resistivity of a high quality ``studied'' layer in which elastic scattering is limited by charged disorder. We find that (i) the additional screening in DLG is relatively inefficient at screening out electron-hole puddles induced by the high concentration of defects in SiO$_2$ as well as defects in close proximity to the ``studied'' graphene layer itself. In the latter cases, the screening provides roughly a factor of 2 suppression within the ``control'' layer density range of $10^{10}-10^{12}cm^{-2}$. Therefore the suppression of electron-hole puddles itself is insufficient to explain the contrast between the strongly insulating behavior observed in DLG~\cite{Geim11} structures and the absence of such in stand-alone graphene samples including highly homogeneous suspended samples~\cite{SuspGeim12}. (ii) The screening effect of ``control'' layer results in a relatively strong suppression of the elastic relaxation rate resulting in gate tunable mobility of the ``studied'' layer in the DLG at intermediate density, $n_s$. (iii) We show that the location of the charged defects limiting the mobility of the DLG can be determined by measuring the ratio of the momentum relaxation and transport relaxation rates. (iv) Additional screening effect of the ``control'' layer strongly suppresses the  potential of Coulomb interaction between charge carriers. (v) Combined effect of (ii) and (iv) results in a strong suppression of the dephasing rate due to inelastic electron-electron collisions which determines the sign and the magnitude of the quantum interference correction to the resistivity. (vi) As a consequence of (ii) and (v) weak localization correction to the resistivity is gate tunable. (vii) The very low concentration of resonant or short range scatterers that has little effect on the Boltzmann part of the resistivity in high quality ``studied'' layer in DLG nevertheless can provide sufficient intervalley scattering for the quantum correction to resistivity to have a negative (weak localization) sign. Moreover, the additional screening in DLG improves the coherence in the system and therefore expands the range in which quantum correction of weak localization sign is observable. (viii) We find that high quality DLG structures where charged defects are the dominant source of elastic scattering manifest a peculiar gate-tunable crossover between weak anti-localization and weak localization magneto-resistance. The latter could be observed in the intermediate density regime in which electron density in the ``studied'' layer is sufficiently high for random inhomogeneity to be neglected and the transport to be metallic and at the same time is low enough for the effect of the ``control'' layer screening to be substantial, see also Sec.~\ref{sec:IntermedDensity}.

\begin{figure}[ht]
\includegraphics[width=1\columnwidth]{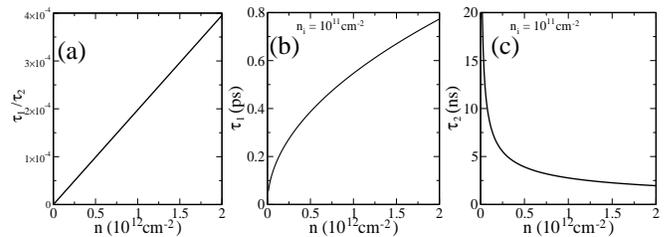} 
\caption{The ratio of intravalley scattering time ($\tau_1$) to intervalley
  scattering time ($\tau_2$), $\tau_1/\tau_2$, or equivalently, the ratio of the
  scattering rates, $\gamma_2/\gamma_1$, where $\gamma_i =
  \hbar/\tau_i$ for the long range Coulomb potential. (b) and (c) show
  $\tau_1$ and $\tau_2$, respectively, as a function of density for an impurity
  density $n_i=10^{11}cm^{-2}$. Note that
  $\tau_1 \propto n^{1/2}$ and $\tau_2 
  \propto n^{-1/2}$.
}
\label{fig:ratio_c}
\end{figure}

In this paper we developed a theoretical framework that can be used for a systematic experimental characterization of the symmetry breaking mechanisms in DLG in the intermediate density regime which could allow further understanding of the observed strongly insulating behavior at low densities. 
This work could also be used to describe a more generic double layer heterostructure involving materials other than graphene that realize other universality classes of Dirac electrons such as surfaces of strong and weak topological insulators. Such heterostructures could be useful to study experimentally gate tunable quantum interference and Anderson localization effects in these systems which are a topic of active current theoretical interest~\cite{LudwigDirac,*Moore,*Stern}.

\section{Aknowledgments}

This work is supported by US-ONR-MURI and NRI-SWAN.

\appendix*

\section{Intervalley scattering due to charged defects}

Intervalley scattering rate due to Coulomb impurities can be separated from the intravalley part by taking 
low and high momentum parts of the sum over $\mathbf{p'}=\pp + \qq + \mathbf{G}$ in Eq.~(\ref{eq:tau}). Where
${\bf G}$ is the reciprocal lattice vector. For intravalley scattering we include only $\mathbf{G}=0$ which results in,
\begin{equation}
\frac{1}{\tau^C_1} = \frac{n_i \nu}{\hbar}
\int_0^{2k_F} \frac{q^2dq}{k_F^3} \left[\frac{V(q)}{\epsilon(q)} \right]^2 \sqrt{1-(q/2k_F)^2} 
\end{equation}
For intervalley scattering we take the smallest allowed reciprocal vector that connects the two corner of hexagonal Brillouin zone, $|{\bf G}|=|\mathbf{K-K'}| = 2\pi/3a$, where $a = 1.42$\AA is the C-C distance. Then the intervalley scattering time is given by
\begin{equation}
\frac{1}{\tau_2^C} = \frac{n_i \nu}{\hbar} \frac{1}{k|{\bf G}-\kk|}
\int_{q_-}^{q_+} q dq [V(q)]^2 \sin \theta,
\end{equation}
where $q_{\pm} = |\GG-\pp| \pm p$ and $\sin \theta$ is given by the following relation 
\begin{equation}
q = \sqrt{|\GG-\pp|^2 + p^2 - 2 |\GG - \pp| p \cos \theta}.
\end{equation}
In Fig.~\ref{fig:ratio_c} we show the calculated ratio of intravalley
scattering time to intervalley scattering time,
$\tau_1^C/\tau_2^C$, as a function of density for the long range Coulomb
impurities. Since the ratio is very small the intervalley scattering
plays a little role in the transport when the Coulomb disorder dominates. 
At the same time a phenomenological model of delta-range disorder gives $\tau_1/\tau_2=1$ at
all densities.



%

\end{document}